# X-ray Properties of Young Early Type Galaxies: II. Abundance Ratio in the Hot ISM

Dong-Woo Kim and Giuseppina Fabbiano

Smithsonian Astrophysical Observatory, 60 Garden Street, Cambridge, MA 02138

Antonio Pipino

ETH Zurich, Wolfgang-Pauli-Str. 27, 8093 Zurich, Switzerland

(Thursday, March 15, 2012)

**ABSTRACT**

Using Chandra X-ray observations of young, post-merger elliptical galaxies, we present X-ray characteristics of age-related observational results, by comparing them with typical old elliptical galaxies in terms of metal abundances in the hot interstellar matter (ISM). While the absolute element abundances may be uncertain because of unknown systematic errors and partly because of the smaller amount of hot gas in young ellipticals, the relative abundance ratios (e.g., the $\alpha$-element to *Fe* ratio, most importantly *Si/Fe* ratio) can be relatively well constrained. We find that in two young elliptical galaxies (NGC 720 and NGC 3923) the *Si* to *Fe* abundance ratio is super-solar (at a 99% significance level), in contrast to typical old elliptical galaxies where the *Si* to *Fe* abundance ratio is close to solar. Also the *O/Mg* ratio is close to solar in the two young elliptical galaxies, as opposed to the sub-solar *O/Mg* ratio reported in old elliptical galaxies. Both features appear to be less significant outside the effective radius (roughly 30'' for the galaxies under study), consistent with the observations that confine to the centermost regions the signatures of recent star formation in elliptical galaxies. Observed differences between young and old elliptical galaxies can be explained by the additional contribution from SNe II ejecta in the former. In young elliptical galaxies, the later star formation associated with recent mergers would have a dual effect, resulting both in galaxy scale winds - and therefore smaller observed amounts of hot ISM - because of the additional SNII heating, and in different metal abundances, because of the additional SNII yields.

*Subject headings*: galaxies: individual (NGC 720, NGC 3923) – galaxies: elliptical and lenticular – galaxies: ISM – X-rays: galaxies – galaxies: abundances

# 1. Introduction

Elliptical galaxies are generally old and passively evolving stellar systems, as their star formation ceased typically several Gyr ago. However, recent measurements have yielded a handful of elliptical galaxies with intermediate stellar age (< 5 Gyr); these galaxies are thought to occupy the evolutionary space between the Antennae-like merging systems and typical old elliptical galaxies (e.g., Thomas et al. 2005; Schweizer 2003). The obvious signatures of young stellar populations, OB stars and type II supernovae (SNe II) are elusive, since they disappear well before a few 100 Myr after the merger-induced star formation episode as the stellar populations evolve. It is the 1-5 Gyr old post-merger system, which carries the key observational evidence of a rejuvenation event. Using a sample of these galaxies (we will refer to them as 'young ellipticals' in this paper) observed with Chandra (Kim and Fabbiano 2010), we have already found a distinctive signature in the X-ray Luminosity Function (XLF) of their low-mass X-ray binary (LMXB) population, which have power-law slopes in between those of typical old elliptical galaxies and of spiral and star burst galaxies with current star formation. Young ellipticals tend to host a relatively larger number of luminous (> 5 x $10^{38}$ erg s$^{-1}$) LMXBs relative to typical old elliptical galaxies.

The star formation history of a galaxy is also reflected in the chemical evolution of its ISM. A key observable to constrain the metal enrichment history is the abundance ratio of α-elements to *Fe*, because of different production yields of SNe Ia and II. Based on the SN yields, it is expected that the *Si* to *Fe* ratio is close to 0.5 solar if SNe Ia are dominating, while ~2.5 solar if SNe II are dominating (e.g., see Kim 2011). Typical old giant elliptical galaxies have *α/Fe* close to solar (e.g., NGC 1399 and NGC 5044, Buote et al. 2002, 2003; N507, Kim & Fabbiano 2004; N7619 Kim et al 2008), indicating that the ejecta from both SNe Ia and II are well mixed. The near-solar *Si/Fe* abundance ratio indicates that ~70% of the detected *Fe* mass was produced by SN Ia. Instead, α elements are significantly enriched in the ISM of merging galaxies with current active star formation (e.g., in the Antennae, Fabbiano et al. 2004; Baldi et al. 2006; NGC 4490, Richings et al 2010). In the recent merger scenario of young elliptical galaxies, we would then expect an enrichment of α-elements from SNe II, associated with the recent star formation (e.g., Kawata & Gibson 2003, Pipino et al. 2005), and enhanced Si/Fe ratios, between 1 and 2.5 solar. In this paper, we intend to test this hypothesis by analyzing Chandra observations for carefully selected young elliptical galaxies.

Young elliptical galaxies tend to have a small amount of hot gas (e.g., Fabbiano and Schweizer 1995; Sansom et al. 2006; Kim et al. in preparation), with $L_X$(gas) comparable to or smaller than $L_X$(LMXB) (see Boroson, Kim, & Fabbiano 2011, hereafter BKF, where the X-ray emission of a sample of elliptical galaxies observed with Chandra is decomposed into the contributions of hot gas, LMXB population, underlying stellar component, and nuclei). Instead, old elliptical galaxies have a wide range of $L_X$(gas), reaching as high as $L_X$(gas) 10 or 100 times higher than $L_X$(LMXB) for group or cluster dominant galaxies. For X-ray faint galaxies (i.e. galaxies with small or no hot ISM, like the young elliptical galaxies), the absolute abundances of individual elements are uncertain leading to controversial claims (see review in Kim 2011). However, we have found that the abundance ratio can be well constrained because of strong correlations between different elements (Appendix; see also Matsushita et al. 2000). Using this effect, we can explore differences in the metal enrichment of old and rejuvenated elliptical galaxies.

This paper is organized as follows. In Section 2, we describe our sample selection, Chandra observations and data reduction techniques. In Section 3, we explain our spectral fitting method, present our results on the abundance ratios from young elliptical galaxies and compare with those of old samples. In Section 4, we discuss the implications of our results on the age effect on the ISM metallicity. In appendix, we briefly describe important technical issues in abundance measurements.

## 2. Sample selection and Chandra observations

To accurately determine the gas abundances and properly compare the abundance ratios between young and old elliptical galaxies, we have selected two young elliptical galaxies with a relatively large amount of hot ISM, NGC 720 and NGC 3923, based on the results of BKF (marked by two big circles in Figure 1a); these two galaxies have stellar ages of 3-5 Gyr (see Table 1). The X-ray luminosity of the hot ISM of these two galaxies is $L_X$(gas) ~ 5 x $10^{40}$ erg s$^{-1}$, comparable to that of the total LMXB contribution. We note that $L_X$(gas) is lower than $L_X$(LMXB) in about 2/3 of the BKF sample and even lower than $L_X$(AB+CV) in about 1/3 of the BKF sample, leading to a large uncertainty in the determination of the X-ray gas properties. There are two other possible candidates for this study in the BKF sample (Figure 1a), NGC 1316 and NGC 4125 with a similar or slightly lower $L_X$(gas), but they are not suitable for our study. The Chandra observations are not deep enough for our purpose, having only net exposure times of 24 and 60 ksec, respectively. With Chandra we can detect luminous LMXBs in both NGC 720 and NGC 3923 and exclude them from the diffuse emission. Fainter, undetected LMXBs contribute about 10-20% of the remaining diffuse emission of these galaxies, and the expected stellar emission from ABs and CVs is less than ~5% of the diffuse emission and therefore it does not significantly affect the gas properties (see Figure 1). Moreover, both NGC 720 and NGC 3923, do not have luminous nuclear sources, therefore there is no potential AGN contamination of the diffuse emission from the extended PSF wing.

We have selected NGC 4472 and NGC 4649, as representative of old ellipticals (age ~ 8-14 Gyr). The X-ray properties of old giant elliptical galaxies have been studied in several galaxies, showing remarkable similarity in terms of abundance ratios (see Kim 2011); NGC 4472 and NGC 4649 have the largest gaseous components in the BKF sample (Figure 1b), ensuring the most significant results of our spectral fits. NGC 4472 and NGC 4649 have a slightly larger amount of the hot ISM than NGC 720 and NGC 3923 (see Figure 1), with $L_X$(gas) ~ $10^{41}$ erg s$^{-1}$, higher than $L_X$(LMXB). Again, the stellar contribution from AB and CV is considerably smaller than that of the hot gas and the AGN does not affect the hot gas emission, because they host only weak nuclei (BKF).

Tal et al. (2009) studied tidal features induced by recent mergers in a complete sample. Three of our sample galaxies are in their sample. Both young galaxies exhibit some signs of recent mergers which include boxy outer isophotes in NGC 720 (tidal parameter, $T_C$ = 0.08) and multiple outer shells in NGC 3923 ($T_C$ = 0.1), while the old galaxy, NGC 4472, has no tidal feature at all ($T_C$ = 0.0).

**Table 1**
Basic properties of sample galaxies

| name | $R_{25}$ (′) | d (Mpc) | scale (kpc/′) | B (mag) | $M_B$ (mag) | K (mag) | log $L_K$ ($L_{K\odot}$) | age (Gyr) | $\sigma_*$ (km/s) |
|---|---|---|---|---|---|---|---|---|---|
| N720  | 2.34x1.20 | 27.67 | 8.0 | 11.13 | -21.08 | 7.271 | 11.30 | 3.4-5.4 | 239 |
| N3923 | 2.94x1.95 | 22.91 | 6.7 | 10.62 | -21.18 | 6.500 | 11.45 | 2.8-3.3 | 268 |
| N4472 | 5.12x4.16 | 16.29 | 4.7 | 9.33  | -21.73 | 5.396 | 11.59 | 7.8-9.6 | 279 |
| N4649 | 3.71x3.01 | 16.83 | 4.9 | 9.70  | -21.43 | 5.739 | 11.48 | 11-14   | 310 |

All four galaxies are classified as E (T=-5) in RC3. In Table 1, we list the basic properties of our sample, including optical sizes in 25[th] mag (from RC3), distances and corresponding scales (from Tonry et al. 2001), B mag (from RC3), and K mag (from 2MASS). The optically measured ages and $\sigma_*$ are taken from Trager et al. (2000), Terlevich & Forbes (2002), Thomas et al. (2005), Howell (2005) and Serra & Oosterloo (2010). We note that the stellar velocity dispersion ($\sigma_*$ = 240 - 310 km s$^{-1}$) and the K-band luminosity (log $L_K$ = 11.3-11.6) between our young and old samples are similar. This is important to separate the effect of stellar abundances, because of the known trends in stellar abundances and abundances ratios as a function of $\sigma_*$ (Pipino 2011; see more in Section 4). The Chandra observation log and total effective exposure time (after screening for background flares) is listed in Table 2.

**Table 2**
Chandra Observations

| Name  | obsids                      | observation date            | total exposure |
|---|---|---|---|
| N720  | 492, 7372, 7062, 8448, 8449 | Oct 12 2000 - Oct 12 2006   | 128 ks |
| N3923 | 1563, 9507                  | Jun 14 2001 - Apr 11 2008   | 93 ks  |
| N4472 | 321                         | Jun 12 2000                 | 32 ks  |
| N4649 | 785                         | Apr 20 2000                 | 24 ks  |

NGC 3923 was observed by our team specifically for the purpose of the present study, while we obtained the Chandra data of the other galaxies from the Chandra archive (http://cxc.harvard.edu/cda/). We used the X-ray data analysis tools available in the **CIAO** package (http://cxc.harvard.edu/ ciao/) and followed the ACIS science threads available in the **CIAO** web site, unless noted otherwise. We reran **acis_process_events** to properly correct for the time-dependent gain and charge transfer inefficiency

(CTI). The latest updates in CALDB (http://cxc.harvard.edu/caldb/index.html) on HRMA (CALDB v 4.1.1) and ACIS contamination (CALDB 4.2) were used in our analysis. We generated light curves to check background flares per each observation and excluded events during the flares (>3$\sigma$; see Kim et al. 2004 for more details). The effective exposure time is reduced by 8-10% due to the flares. The X-ray spectra were extracted from different radial bins, excluding all detected point sources using a circle with a radius of 2.5", or a radius corresponding to 95% encircled energy fraction at 1.5 keV, if larger than 2.5". Because these galaxies do not have strong nuclear emission, we used the same radius to exclude the nuclear sources. The response files, rmf (response matrix file) and arf (ancillary reference file), were generated for each source region.  For data taken in multiple exposures, to take into account the ACIS response degradation due to the filter contamination, we generate arf per individual observation and then took an exposure-weighted mean by applying ***dmarfadd*** (for weighted sum) and ***dmtcalc*** (to divide by the number of observations).  The background spectra were extracted from source free regions within the same CCD.

## 3. Spectral Fitting and Abundance Ratios

Excluding the contribution from detected point sources (AGNs and LMXBs) and subtracting the background emission, the remaining diffuse emission will be composed of both hot gas emission and undetected stellar sources.  We assumed a 7 keV Bremsstrahlung to model undetected LMXBs (see Irwin et al. 2003; BFK). We did not consider ABs and CVs, because their contribution is negligible. Following BKF, we tested the AB + CV contribution using a composite APEC + power-law model; we found that our results do not change significantly, except that the normalization of the LMXB component (the 7 keV Bremsstrahlung) slightly decreases due to the hard power-law component of the AB + CV model.

We modeled the hot ISM emission with the APEC model, which includes the up-to-date atomic data available in ATOMDB v2.0.1 (http://www.atomdb.org/). Using these improved atomic data we find that the total $\chi^2$ is reduced by ~10%, when compared to the fitting result with the former atomic data (ATOMDB v.1).  It also changes the gas temperature by ~10% and Fe abundance by ~20%. We allowed all elements to vary independently in the fit within 0.1 and 5 times solar. We discuss in Appendix the problem associated with arbitrarily tying elements, which has been a common practice in the X-ray spectral fits of low signal-to-noise ratio data. Given the temperature (0.5-0.6 keV) of the hot ISM of the two young ellipticals, we measure *O, Mg, and Si* relative to *Fe.* Other elements (e.g., S) which are peaking at higher gas temperatures are not well constrained. We note that while the absolute abundance is critically subject to systematic uncertainties, the relative abundance is well determined. We further discuss the systematic errors in the Appendix. We also note that the error of the abundance ratio cannot be determined simply by adding in quadrature the errors of the individual elements, because these errors are strongly correlated, as shown by the elongated confidence contours. To measure the error of the abundance ratio, we used a dummy model with a free parameter linked to the abundance ratio (but with zero normalization) and determined the error of that parameter (see also

Richings et al 2010). We adopted Grevesse & Sauval (1998) as the solar abundance standard. Throughout this paper, the quoted error is at the 1σ level, unless noted otherwise.

Table 3   Spectral fitting results of two young elliptical galaxies

```
--------------------------------------------------------------------------------
NGC 720
                    r=0-30                 r=30-60                 r=60-120
               7866.25 / 8592          5997.09 / 8422          4157.49 / 9348
--------------------------------------------------------------------------------
2D
χ²red (χ²/dof) 0.73 (73.14/100)       1.05 (122.27/116)       0.87 (133.87/154)
          kT:  0.56 (-0.02 +0.02)     0.57 (-0.02 +0.02)      0.58 (-0.04 +0.03)
       Si_Fe:  1.88 (-0.35 +0.37)     1.26 (-0.42 +0.43)      1.84 (-1.87 +1.47)
        O_Fe:  0.97 (-0.20 +0.22)     0.60 (-0.21 +0.25)      0.35 (-0.92 +0.56)
       Mg_Fe:  1.28 (-0.18 +0.19)     1.22 (-0.21 +0.22)      0.94 (-0.71 +0.62)
        O_Mg:  0.76 (-0.17 +0.21)     0.49 (-0.21 +0.22)      0.37 (-1.18 +1.05)
3D
χ²red (χ²/dof) 0.73 (73.14/100)       1.01 (124.05/116)       0.87  (133.76/154)
          kT:  0.56 (-0.02 +0.03)     0.57 (-0.03 +0.02)      0.55 -(0.03 +0.03)
       Si_Fe:  2.20 (-0.63 +0.57)     0.96 (-0.46 +0.77)      1.83 (-0.94 +1.03)
        O_Fe:  1.11 (-0.29 +0.30)     0.52 (-0.21 +0.26)      0.95 (-0.31 +0.27)
       Mg_Fe:  1.30 (-0.26 +0.25)     1.34 (-0.26 +0.26)      0.66 (-0.30 +0.32)
        O_Mg:  0.80 (-0.28 +0.42)     0.54 (-0.22 +0.24)      1.39 (-0.76 +1.71)
--------------------------------------------------------------------------------
NGC 3923
                    r=0-30                 r=30-60                 r=60-120
               7638.34 / 8088          2235.41 / 3695          1812.55 / 7745
--------------------------------------------------------------------------------
2D
χ²red (χ²/dof) 0.97 (84.22/87)        1.03 (77.97/76)         0.82 (133.76/163)
          kT:  0.47 (-0.02 +0.02)     0.57 (-0.04 +0.02)      0.58 (-0.03 +0.03)
       Si_Fe:  1.41 (-0.30 +0.32)     1.47 (-0.64 +0.68)      1.80 (-1.79 +1.49)
        O_Fe:  0.72 (-0.13 +0.14)     1.52 (-0.39 +0.45)      0.53 (-0.64 +0.75)
       Mg_Fe:  0.80 (-0.13 +0.13)     0.50 (-0.28 +0.27)      1.06 (-0.73 +0.58)
        O_Mg:  0.91 (-0.21 +0.27)     3.06 (-0.99 +2.28)      0.50 (-0.61 +1.46)
3D
χ²red (χ²/dof) 0.97 (84.18/87)        1.08 (82.32/76)         0.82 (133.76/163)
          kT:  0.46 (-0.02 +0.02)     0.56 (-0.05 +0.04)      0.58 (-0.03 +0.04)
       Si_Fe:  1.42 (-0.38 +0.37)     1.53 (-1.04 +1.19)      1.77 (-1.50 +1.51)
        O_Fe:  0.64 (-0.12 +0.13)     1.96 (-0.49 +0.76)      0.48 (-0.57 +0.77)
       Mg_Fe:  0.83 (-0.15 +0.16)     0.43 (-0.42 +0.38)      1.04 (-0.66 +0.58)
        O_Mg:  0.72 (-0.16 +0.20)     5.45 (-1.53 +3.67)      0.52 (-0.59 +0.82)
--------------------------------------------------------------------------------
```

The results of the spectral fitting are given in Table 3 for spectra extracted from three radial bins (0-30", 30-60" and 60-120"). The outer radius of the first bin is particularly selected such that this region can have enough net counts, but without mixing different gas in different temperatures and possibly in different abundances (see below for more). The net and total counts, listed in the 3rd line of the header, indicate that the background emission is almost negligible in the central bin and increases progressively toward the outer bins. In both NGC 720 and NGC 3923, the gas temperature does not change significantly. This ensures that we can model the gas emission from each bin with only one APEC component. We also corrected for the projection effect (listed under 3D in Table 3) by applying the onion peeling technique, ***deproject***, available in the CIAO contributed software (http://cxc.harvard.edu/contrib/deproject/); we found that our results do not change significantly.

The fits are all reasonably good, with reduced $\chi^2$ slightly less than or close to 1 in all regions. Among $\alpha$-elements, the *Si* abundance is the most constrained, because of the strong, isolated *Si* XIII emission feature at E ~ 2 keV (see discussion in Kim 2011). The theoretical yield of *Si* is also well determined, with the least amount of scatter between model predictions (e.g., Gibson et al. 1997; Nagataki & Sato 1998). In the central region (r < 30") of NGC 720, the *Si/Fe* ratio is 1.88 (-0.35 +0.37) solar, super-solar at the 99% significance level. In Figure 2, we show the 39%, 68% and 90% confidence contours of *Si* and *Fe* abundances. The super-solar *Si/Fe* ratio is clearly seen. When the projection effect is corrected (under 3D in Table 3), the *Si/Fe* ratio increases to 2.2 solar, because of the negative radial gradient of the *Si/Fe* ratio but the corresponding error increases because of the dependency on the results of the outer regions. The significance of the super-solar ratio remains similar at a 99% level.

In the central region of NGC 3923, the best-fit *Si/Fe* ratio is again super-solar (1.4 solar) in both 2D and 3D, but the significance of the super-solar ratio is only marginal (slightly over 1$\sigma$ level). See also Figure 2b. This is partly because NGC 3923 has a smaller amount of hot gas than NGC 720, its ISM is slighter cooler and the Galactic line of sight absorption is higher.

We applied the same technique to the old elliptical galaxies NGC 4472 and NGC 4649 (see Table 4). Both of these galaxies host no strong AGN and no obvious gas structure as our two young elliptical galaxies. However, they contain a larger amount of hot gas than young elliptical galaxies. Consequently they are more extended and exhibit a positive gas temperature gradient, as they are embedded in the Virgo cluster (i.e., inside the hotter ICM). Here we analyze the spectra extracted only from the central region where the effect of the hotter ICM (as well as the background) is negligible. As before, we select r = 30" in an angular size. If we use the same physical size for the extraction region (~50% larger than that of young elliptical galaxies), the results do not change significantly, but the reduced $\chi^2$ increases by 10-20%, indicating that a single temperature model becomes less applicable. With our selection of r=30", the goodness of the fit is excellent with reduced $\chi^2$ slightly less than or close to 1 in both old elliptical galaxies. In contrast to the two young elliptical galaxies, the *Si/Fe* ratios of both old elliptical galaxies are remarkably close to the solar ratio (0.9 - 1 solar with an error of 0.1 solar). This is consistent with earlier results in similar old elliptical galaxies (see section 4). In Figure 2 (c and d), we also show the confidence contours of these old elliptical galaxies. The contrast between young and old elliptical galaxies is quite remarkable.

To further visualize the difference, we directly compare the *Si/Fe* ratio in Figure 3. To improve statistics, we also consider other relaxed old elliptical galaxies from the literature. The two other galaxies with high $L_X$(gas) in Figure 1b (NGC 4374 and NGC 4261) are not good candidates because both of them exhibit complex gas sub-structures, possibly associated with central radio sources (3C272.1 and 3C270, respectively), which make spectral results less reliable. We take the results from Humphrey and Buote (2006) who analyzed Chandra spectra with most important elements allowed to vary freely and published the errors of abundance ratios. Among their first group of high $L_X$ galaxies, we use four relaxed old elliptical galaxies (NGC 507, NGC 1399, NGC 5846 and NGC 7619) with accurate abundance ratio measurements (the error of Si/Fe less than 30%). Due to the different solar standard, we convert their abundance ratios to GRSA (most significantly raising O by a factor of 1.32.) We do not use the other galaxies in the same group because they are either affected by AGN or not old (or no age measurement), or their abundance ratios are subject to large errors. In Figure 3, we plot the average *Si/Fe* ratios in two young (the big red square) and six old (the big blue square) elliptical galaxies, which clearly visualize the difference. We note that Humphrey and Buote (2006) also analyzed the earlier Chandra data of NGC 720 (29ks) and NGC 3923 (9ks), but due to the short exposures, they could not measure Si/Fe reliably (no measurement in NGC 720 and a lower limit of 2.1 solar in NGC 3923).

Table 4  Spectral fitting results of the central region (r<30") of two old elliptical galaxies

|  | NGC 4472 | NGC 4649 |
|---|---|---|
| 2D |  |  |
| $\chi^2_{red}$ ($\chi^2$/dof) | 0.92 (99.90/109) | 1.05 (108.35/103) |
| kT | 0.83 (-0.01 +0.01) | 0.86 (-0.01 +0.01) |
| Si_Fe | 1.03 (-0.11 +0.11) | 0.90 (-0.11 +0.10) |
| O_Fe | 0.87 (-0.15 +0.10) | 0.40 (-0.14 +0.09) |
| Mg_Fe | 1.43 (-0.12 +0.12) | 1.34 (-0.13 +0.12) |
| O_Mg | 0.61 (-0.12 +0.09) | 0.30 (-0.10 +0.08) |

We also measure the *O/Fe* and *Mg/Fe* abundance ratios (see table 3 and table 4). The *O/Mg* abundance ratio is particularly interesting, because both *O* and *Mg* are mainly produced by SN II. Based on the SN II yield alone, the O/Mg ratio should be close to solar (see Kim 2011). However, sub-solar *O/Fe* and *O/Mg* ratios (~0.5 solar) have been reported for typical old elliptical galaxies, spurring discussion of the cause of the reduced *O* abundance (e.g., Buote et al. 2003; Kim & Fabbiano 2004; David et al. 2011). Here, we find that the *O/Mg* ratios are close to the expected solar in the central regions of the two young elliptical galaxies: *O/Mg* = 0.8 (-0.2, +0.3) solar and 0.9 (-0.2, +0.3) solar for NGC 720 and NGC 3923, respectively. In contrast, in our two old elliptical galaxies the *O/Mg* ratio is 0.3-0.6 solar (with an error of 0.1-0.2 solar), consistent with previous reports for similar old elliptical galaxies. In Figure 3, we compare the abundance ratios (*O/Fe, Mg/Fe* and *O/Mg)* in young and old elliptical galaxies. It is clear that while the *O/Fe* and *Mg/Fe* ratios are consistent with each other within errors in young elliptical galaxies, the *O/Fe* ratio is considerably lower than the *Mg/Fe* ratio in old elliptical galaxies. We will further discuss the implication of the age-dependent *O/Mg* ratio in section 4.

The recent star formation episode is often restricted to the circum-nuclear regions (e.g., Kuntschner et al. 2010; Koleva et al. 2011; see more discussions in section 4). With the X-ray spectra extracted from r= 30"-60" we can investigate whether the age-abundance effect is also limited to the central region. We note that in both NGC 720 and NGC 3923, the temperature gradient is not significant. In NGC 720, the gas temperature remains the same, kT = 0.56 - 0.58 keV in r < 120". In NGC 3923, the temperature is kT = 0.47 keV in r < 30" and slightly increases to kT = 0.57 keV in r = 30-120". Therefore, we can still apply one-component APEC models to represent the gas emission in the r=30-60" annulus, as in the central region (r < 30"). In both young elliptical galaxies, the two distinct features seen in the center, super-solar *Si/Fe* and solar *O/Mg* ratios, are not found at larger radii. The *Si/Fe* ratio is 1.26 ± 0.4 and 1.47 (-0.6, +0.7) solar in NGC 720 and NGC 3923, respectively, fully consistent with the solar ratio. Similarly, the *O/Mg* ratio is 0.5 ± 0.2 solar in NGC 720, consistent with those of old elliptical galaxies; this ratio is not well determined in NGC 3923. In Table 3, we list the result in the 3$^{rd}$ bin at r=60-120" for completeness, but neither *Si/Fe* nor *O/Mg* can be determined. The ISM abundance ratios specific for the younger generation of stars seem to be confined to the central region in young elliptical galaxies. We will further discuss the implication of the localized age effect in section 4.

Ji et al. (2009) studied both NGC 720 and NGC 3923 with the XMM-Newton RGS spectra, but they could not measure *Mg* and *Si* due to the limited high energy response of RGS. In their analysis with XMM-Newton EPIC data, the smallest used region is r < 1', which is too large to identify the distinct features we have found in our Chandra analysis of these two young elliptical galaxies. In NGC 4472 and NGC 4649, Ji et al's results with Chandra ACIS and XMM-Newton EPIC spectra (two-phase model fitting to spectra extracted from a larger area) are consistent with ours, i.e., near-solar Si/Fe and sub-solar O/Mg ratios. Similarly, analyzing Suzaku data of NGC 4472, Loewenstein & Davis (2010) reported near-solar *Si/Fe* and *Mg/Fe* ratios and half solar *O/Mg* ratio both in the inner (r < 4') and outer (r = 4'-8') regions.

## 4. Discussion

We have studied the spectral behaviour of the hot ISM of the two young elliptical galaxies NGC 720 and NGC 3923, which, among those of their kind observed with Chandra, are the best candidates for accurate measurements of abundance ratios. These two galaxies have the largest amount of hot gas among nearby young elliptical galaxies with deep Chandra observations, weak or no AGN emission and no radio jet. Also they do not have obvious gas structural features, such as filaments, tails, cavities and clumps which could make the X-ray spectral analysis harder due to the complex structure of temperature and abundance. In these galaxies we find abundance ratios different from those found in typical old elliptical galaxies. The *Si* to *Fe* abundance ratio is super-solar in NGC 720 and NGC 3923, contrary to the close to solar *Si/Fe* reported in old elliptical galaxies (e.g., Buote et al. 2002, 2003; Kim & Fabbiano 2004; Humphrey and Buote 2006; Kim et al. 2008; Ji et al. 2009; David 2011) and confirmed by our analysis of NGC 4472 and NGC 4649. The *O/Mg* ratio is close to solar, as opposed to the sub-solar *O/Mg* ratio reported in old elliptical galaxies (e.g., Buote et al. 2002, 2003; Kim & Fabbiano 2004; Humphrey & Buote 2006; Ji et al. 2009) and also confirmed in this study. Both features appear to be significant only in the central region (r < 30").

The *Si/Fe* abundance ratio is expected to be ~2.5 solar, if heavy elements were mainly synthesized in SN II; for example, significantly enhanced α to *Fe* ratios are seen in metal poor Galactic halo stars and local group dwarf galaxies with low *Fe/H* ratios which basically represent pure SN II products (e.g., see Pagel 2009). This ratio would decrease with increasing contributions from SN Ia. If the SN Ia contributed most of the hot ISM metals, the *Si/Fe* abundance ratio would be ~0.5 solar (e.g., Iwamoto et al. 1999). Typical well-studied elliptical galaxies have the *Si/Fe* ratio close to solar. Examples are NGC 1399 (Buote 2002), NGC 5044 (Buote et al. 2003), NGC 507 (Kim & Fabbiano 2004), NGC 7619 (Kim et al 2008) and NGC 4472 (Loewenstein & Davis 2010); these are all old giant elliptical galaxies with large amounts of hot ISM. Additional examples can be found in the Chandra and XMM-Newton archival studies by Humphrey and Buote (2006) and Ji et al (2009). The solar *Si/Fe* ratio indicates that the ejecta from both SNe II (from early star formation) and SNe Ia (continuously added later) are well mixed in these old elliptical galaxies. With SN yields taken from Iwamoto et al. (1999), the solar abundance ratio of *Si* to *Fe* indicates that approximately 70% of the *Fe* mass was produced by SN Ia. In most cases, the *Si/Fe* abundance ratio does not show any measureable radial gradient (e.g., NGC 507 by Kim & Fabbiano 2004; NGC 4472 by Loewenstein & Davis 2010), while the individual abundance (e.g., *Fe* and *Si*) decreases with increasing radius. This uniform *Si/Fe* ratio further suggests that in typical old elliptical galaxies, SN II and SN Ia ejecta are well mixed on a larger scale (~100 kpc) than the optical galaxy (~40 kpc).

In the central region of the young elliptical galaxies NGC 720 and NGC 3923 instead, we find that the *Si/Fe* ratio is super-solar (1.4 -1.9 solar), clearly indicating additional contribution from SN II ejecta, when compared to the old elliptical galaxies. In these young elliptical galaxies, a considerable amount of the hot ISM must have been lost due to the extra energy from the later star formation possibly associated with recent mergers and more α elements (relative to *Fe*) from new SNe II ejecta were added to the hot ISM. A simple estimate with SN yields from Iwamoto et al. (1999) suggests that *Fe* (in mass) comes from SN Ia and SN II roughly in equal proportions.

In a simple scenario the entire star formation history can be modeled with two starburst episodes: the first generation which is similar to that of old ellipticals and the second generation which occurs at later time due to a single major merger or series of minor mergers. If all the SNe ejecta were retained (i.e., in a closed box model), the observed *Si/Fe* ratio of ~1.5 solar would require that about half of the stars belong to the second generation. This constraint can be relaxed if a considerable amount of the hot ISM is lost during the period of the second star formation episode, as suggested by observational results (e.g., Fabbiano & Schweizer 1995; Kim et al. in preparation). For example, if only 10-20% of the original hot ISM survived through the recent mergers, the observed *Si/Fe* ratio requires that roughly 10-20% of stars may belong to the second generation. This is consistent with the observed fraction in the central region of young elliptical galaxies. For example, Annibali et al. (2007) estimated that among young nearby early type galaxies the rejuvenation episode involves < ~25% of the total galaxy mass (see also Serra & Oosterloo 2010).

In order to illustrate the chemical enrichment history of the hot ISM, we show in Figure 4a the *Si/Fe* abundance ratios predicted by chemical evolution models for young elliptical galaxies with additional star formation. The "case 2a" model discussed in Pipino & Matteucci (2011) is marked by a black solid line. This model successfully reproduces both the stellar and the gaseous element pattern of typical old elliptical galaxies. The secondary star formation ignited arbitrarily at t = 9 Gyr for 1 Gyr, resulting in ~25% of the final mass in the 2$^{nd}$ generation stars. The star formation intensity and duration are chosen so that such a secondary burst can be considered a rejuvenation episode in line with the above observational estimates. Since the secondary burst will make the predicted SN Ia rate higher than the observed one (red dotted in Figure 4a), we further assume that the effective probability of binaries becoming SN Ia is reduced at t > 9 Gyr by a factor of 2-3 with respect to the standard case to match with the observational present-day estimate in clusters (blue dotted) and in the field (green dotted) (see Mannucci et al. 2008 and Fig.1 in Pipino & Matteucci 2011). We also show the prediction (blue dashed) with up-to-date SN Ia yields taken from the C-DDT model which incorporates more accurate 2D SN explosion simulations (Maeda et al. 2010). This corresponds to "case 3a" of Pipino & Matteucci (2011). Also plotted in Figure 4a are the observed *Si/Fe* ranges for NGC 720 (red shaded) and NGC 4472 (blue shaded) for comparison. In all cases, the rapid injection of the additional SN II ejecta makes the *Si/Fe* ratio jump to ~2.5 solar at the onset of the star burst. Then the *Si/Fe* ratio slowly decreases and reaches at 1.2-2 solar at 3 Gyr after the burst. Our observed super-solar *Si/Fe* ratios (1.4-1.9 solar) in two young elliptical galaxies are well within the range predicted by the chemical evolution models. The single stellar population (SSP) equivalent age of a few Gyr may actually mean that a smaller fraction (~5%) of their stellar mass is contained in an even younger population formed during the past < ~1Gyr (e.g., Serra & Oosterloo 2010), because $t_{SSP}$ is strongly biased towards the age of the youngest stars (see Serra & Trager 2007). In this case, the peak in Figure 4a would shift downward and toward the later time, still consistent with the observed super-solar *Si/Fe* ratio. In this model calculation, we did not consider several potentially important effects, such as the possibly different IMF during the secondary burst, metallicity effects on SN yields etc., which are beyond the scope of this paper. A comprehensive study with more realistic simulations of bursts is left to a forthcoming paper.

The *O* abundance is sub-solar in most previous studies of old elliptical galaxies. The *O* to *Fe* ratio is ~0.3 solar in NGC 5044, (Buote et al. 2003; David et al. 2011), 0.3-0.5 solar in NGC 507 (Kim & Fabbiano 2004) while the *Mg* to *Fe* ratio is only slightly lower than solar (0.8-0.9 solar) in both galaxies. Recent studies with Chandra and XMM-Newton archival data by Humphrey and Buote (2006) and Ji et al (2009) and with Suzaku data of NGC 4472 by Loewenstein & Davis (2010) also show the same trend: the *O/Fe* ratio is ~0.5 solar, while the *Mg/Fe* ratio is close to solar, hence the *O/Mg* ratio ~0.5 solar. The above observed ratios were based on the solar abundance standard given by Grevesse & Sauval (1998), except for those of Humphrey and Buote (2006) who adopted an older version of Asplund et al. (2009). Both ratios would be higher by a factor of 1.4 when converted from the former standard to the latter due to the lower *O* abundance in the latter (e.g., see Kim 2011). With a simple combination of SN II and SN Ia ejecta, for example, if the *Si/Fe* ratio is close to solar (as found in old elliptical galaxies), the *O/Fe* ratio would be 0.9-1.3 solar depending on the adopted solar standard. Since both *O* and *Mg* are mainly produced by SN II, the *O/Mg* ratio does not significantly depend on the SN Ia rate. Based on the SN II yield alone, the *O/Mg* ratio should be close to solar, regardless of the *Si/Fe* ratio. The observed sub-solar *O/Fe* and *O/Mg* ratios are inconsistent with those expected from the SN yields, and prompted several investigators to suggest various possible explanations including a warm absorber (Buote et al. 2003), Population III hypernovae (Humphrey & Buote 2006) and incorrect standard core collapse nucleosynthesis models which simply overproduce *O* by ~2 (Ji et al 2009; Loewenstein & Davis 2010).

In the young elliptical galaxies NGC 720 and NGC 3923, we find near-solar *O/Mg* abundance ratios which are close to what is expected from SN II. We must be witnessing the pure (or dominant) SN II yields in these young elliptical galaxies where the previously accumulated ISM might have been expelled by merger induced galactic winds. However, quantitatively the different *O/Mg* ratio between young and old elliptical galaxies and the sub-solar *O/Mg* ratio in old elliptical galaxies are still puzzling, because both *O* and *Mg* are predominantly produced by SN II. If both results are correct, the SN II yields ought to be altered in a sense that more *O* must be produced relative to *Mg* in SNe II associated with the recent star formation episode than those associated in the early star formation, implying time dependent relative yields. Because the yield critically depends on the mass and metallicity of progenitors (e.g., Nomoto et al. 2006; Kobayashi et al. 2006), the old and new generation of stars might have different IMF and/or metallicity. For example, more massive, low metallicity progenitors tend to produce more *O* relative to *Mg*. However, the detailed dependency is yet to be understood.

Alternatively, if the early galactic wind, typically starting within 1-2 Gyr from the initial star burst, was able to remove the entire ISM, it could effectively erase the SN II signature in the ISM in ~2Gyr. In that case, the SN II signature only remains in stars which continuously inject mass loss into the ISM, as the stellar system evolves passively. Although SNe Ia do not produce *O* and *Mg* as much as SNe II, SNe Ia yields become important in determining the *O/Mg* ratio in the ISM. In Figure 4b, we show the *O/Mg* ratios predicted by the wind model with and without the 2$^{nd}$ star formation at t = 9 Gyr. As discussed in Pipino & Matteucci (2011), the base model (case 2a) cannot reproduce the sub-solar *O/Mg* ratio observed in the typical old elliptical galaxies (black solid line). However, the *O/Mg* ratio can be considerably reduced in case 3a (with the new SN Ia yields from the C-DDT model, Maeda et al. 2010) or case 1c where the *Mg* yield is arbitrarily increased by a factor of five, to match that empirically determined with abundances in the Milky Way by Francois et al. (2004). In the C-DDT model, a single SN Ia explosion produces 0.34 M$_\odot$ of *O* and 0.059 M$_\odot$ of *Mg* which make the *O/Mg* ratio a factor of 2 lower than that predicted by the previous standard models (e.g., WDD1 in Iwamoto et al. 1999). In this case, the *O/Mg* ratio is sub-solar due to the reduced *O/Mg* ratio in SN Ia yields (blue solid line in Figure 4b). With the 2$^{nd}$ star burst, while 'case 2a' does not significantly change the *O/Mg* ratio (green dotted), the SN II ejecta in 'case 3a' raises the *O/Mg* ratio, slightly higher than solar (blue dashed line). Quantitatively, the case 3a predictions are still higher than observed O/Mg ratios in both old and young elliptical galaxies. However, this model shows one example of how to create the difference in the *O/Mg*. Again, more observational data and comprehensive considerations of various important factors are necessary before reaching a definitive conclusion for the chemical evolution of the hot ISM in young elliptical galaxies. However, it is clear that our pilot study of young elliptical galaxies does open up a new path toward better understanding of the galaxy evolution and calls for more works both in theory and observation.

Both the super-solar *Si/Fe* and solar *O/Mg* of the young elliptical galaxies are significant only in the central region, r < 30". This angular size corresponds approximately to one effective radius, r$_e$ = 25" and 40" in NGC 720 and NGC 3923, respectively (from 2MASS Large Galaxy Atlas). In the outer region (r = 30-60"), both abundance ratios are consistent with those of the old galaxies, within the statistical errors. In young early type galaxies, the recent star formation episode is often restricted to the circum-nuclear regions, sometimes associated with dusty disks/rings (e.g., Kuntschner et al. 2010; Koleva et al. 2011). While some young galaxies show a spatially extended recent star formation episode, the youngest stellar populations are still found in the central parts (Kuntschner et al. 2010). Our results are consistent in that the ISM abundance ratios are revealing the recent star formation where it is most prominent.

We note that the stellar velocity dispersion and the K-band luminosity of our young elliptical galaxies are similar to, but slightly lower than those of old elliptical galaxies (see Table 1). It is important to separate the effect of stellar abundances on the ISM abundances, because the mass loss of evolved stars is an important source of the ISM and because the stellar abundances depend on $\sigma_*$. It is known that the stellar $\alpha$-elements are significantly enhanced (relative to *Fe*) in old giant elliptical galaxies (e.g., Thomas et al. 2005, see also Pipino 2011). This is often explained by the shorter star formation time scale for bigger galaxies (so called downsizing) such that these galaxies did not have enough time to add *Fe* from SN Ia ejecta to stars. This effect is opposite from what we have shown in section 3, where the *Si* abundance (relative to *Fe*) is enhanced in the ISM of less massive younger galaxies, therefore our results are not affected by the different stellar abundance.

The data analysis was supported by the CXC CIAO software and CALDB. We have used the NASA NED and ADS facilities, and have extracted archival data from the Chandra archives. This work was supported by the Chandra GO grant G08–9133X (PI: Kim), XMM-Newton GO grant NNX09AT20G (PI: Kim) and NASA contract NAS8-39073 (CXC).

# Appendix

## Technical Issues in Abundance Measurements by X-ray Spectral Fitting

Ideally, metal abundances can be measured by fitting proper models to the observed X-ray spectra and the related uncertainties can be constrained by applying proper statistics. Practically, however, there are various systematic effects and simplified assumptions which affect the results. See Kim 2011 for in-depth discussions. Here we describe a few issues related to our results.

We would like to emphasize that while uncertainties associated in the absolute abundance of individual elements could be quite larger than that given by the fitting statistics, the relative abundances or abundance ratios can still be relatively reliable, even if some systematic error are not fully considered. This is because the abundances (and the errors) of different elements are strongly correlated, as reflected by the diagonally elongated confidence contours in Figure 2 (see also Richings et al. 2010). The strong correlation among different elements is because the uncertainty in one element affects the other element in a similar way, i.e., if one element is underestimated due to a given uncertainty, another element may also be underestimated for the same reason. For example, if the continuum level increases by underestimating the background emission or undetected LMXB contribution, both *Fe* and *Si* will be underestimated in a similar manner. Uncertainties in abundance measurements may also be caused by inappropriate model selections (e.g., single components vs. multiple components, MEKAL vs. APEC) and other unknown systematic effects (e.g., *He* sedimentation). However, the relative abundances may still be reliable.

To illustrate the robustness of the abundance ratio, even with the erroneous absolute abundances, first we compare the two most popular thermal gas models, APEC and MEKAL. In the case of the central region of NGC 720, the best fit values of *Fe* and *Si* are 1.0 (±0.2) solar and 1.8 (±0.7) solar for the MEKAL model and 0.68 (±0.2) and 1.31 (±0.4) solar for the APEC model. Not only the overall statistics (in $\chi^2$) is improved by ~15% in the APEC model, but also the local deviations (localized excess and deficit in $\Delta\chi$) in the MEKAL model disappears in the APEC model. With more accurate plasma emission data (ATOMDB v2), APEC seems to better represent the hot plasma emission than MEKAL. It is interesting to note that the *Si/Fe* ratio is very similar (1.9 solar) in both models (regardless of their validity), even though the absolute abundances are different. Although it is not clear why MEKAL returns high absolute abundances, MEKAL seems to produce a lower continuum level in the above example. However this may vary case by case.

As another example, we consider the effect of the *He* abundance. The Helium abundance is usually assumed to be solar. However, non-solar *He* abundance can affect other element abundances by changing the continuum emission, e.g., by helium sedimentation (e.g., Fabian & Pringle 1977; Gilfanov & Syunyaev 1984). Considering relaxed galaxy clusters, Ettori & Fabian (2006) showed that a He enhancement by a factor of a few with respect to the solar value could reduce by ~50% the metallicity measured in the center of the ICM and might explain the metallicity drop reported in the inner (~20 kpc) regions of nearby bright galaxy clusters (e.g. Centaurus, Sanders & Fabian 2002; A2199, Johnstone et al. 2002). If we set the *He* abundance to be 5 times solar in spectral fitting in the central region of NGC 720, the *Fe* abundance becomes double. Similarly the *Si* abundance is also double, making the *Si/Fe* ratio almost identical.

Another important issue in X-ray spectral fitting is that various element abundances are often tied at the solar ratio. This practice has been customary mainly because the observed spectra do not have enough statistics to allow many degrees of freedom. Matsushita et al. (2000) pointed out that the Fe abundance is significantly different, when measured with solar and non-solar $\alpha$-element to *Fe* ratios. David et al. (2011) further demonstrated (see their Fig. 5.7) that tying *O* to *Fe* results in an incorrect *Fe* abundance by artificially altering the continuum level at low energies (~ 0:5 keV). See Figure 6 in Kim 2011 for more discussion. We do not have a priori information about how they can tie with each other because some elements are expected to deviate from the solar abundance, e.g., by different relative contribution from different types of SN ejecta or by mixing different gas by merger, infall etc. Furthermore, even if *O* and *Mg* abundances are supposed to be similar (i.e., *O/Mg* ~ solar) because both are mainly produced by the same mechanism, they are quite different in old elliptical galaxies (see section 5). Therefore the only reliable method is to set all elements vary independently, except a few elements which have negligibly weak emission features. The situation becomes more complicated when the ISM consists of multi-phases with multiple temperatures. For example, when two-component thermal gas models are applied to represent the gas in the temperature gradient, the abundances in two thermal models are likely different. However, we note that in almost all previous studies, the abundances in two components are tied, simply because they cannot be measured separately.

**Reference**

Annibali, F., A. Bressan, R. Rampazzo, W. W. Zeilinger, and L. Danese. 2007 A&A 463, 455
Arimoto, N., Matsushita, K., Ishimaru, Y., Ohashi, T., & Renzini, A. 1997, ApJ, 477, 128
Asplund, M., Grevesse, N., Sauval, A. J., & Scott, P. ARA&A 47, 481 (2009)
Baldi, A. et al. 2006, ApJ, 636, 158
Boroson, B., Kim, D.-W., & Fabbiano, G. 2011, ApJ, 729, 12 (**BKF**)
Buote, D. A., & Fabian, A. C. 1998, MNRAS, 296, 977
Buote, D. A. 2002, ApJ, 574, L135
Buote, D. A., Lewis, A, D., Brighenti, F., & Mathews, W. G. 2003, ApJ, 595, 151
David, L. P. et al. 2011 ApJ, 728, 162
Ettori, S. & Fabian, A. C. 2006, MNRAS, 369, L42
Fabbiano, G. 1995 *Fresh Views of Elliptical Galaxies*, ed. A. Buzzoni et al. (San Francisco: ASP) p103
Fabbiano, G. & Schweizer, F. 1995, ApJ, 447, 572
Fabbiano et al. 2004, ApJ, 605, L21
Fabian, A. C., Pringle, J. E., 1977, MNRAS, 181, 5
Francois, P., et al. 2004, AA, 421, 613
Gibson, B. K., Loewenstein, M., & Mushotzky, R. F. 1997, MN, 290, 623
Gilfanov M. R., Syunyaev R. A., 1984, Soviet Astron. Lett., 10, 137
Grevesse, N. & Sauval, A.J. 1998, Space Science Reviews 85, 161
Howell, J. H. 2005 AJ, 130, 2065
Humphrey, P. J., & Buote, D. A. 2006, ApJ, 639, 136
Iwamoto, K., et al. 1999, ApJS, 125, 439
Ji, J. et al. 2009, ApJ, 696 2252
Johnstone, R. M. Allen, S. W. Fabian, A. C. & Sanders, J. S. 2002 MNRAS 336 299
Kawata, D. & Gibson, B. K. 2003, MNRAS, 346, 135
Kim, D.-W., Fabbiano, G., Matsumoto, H., Koyama, K., & Trinchieri, G. 1996, ApJ, 468, 175
Kim, D.-W., & Fabbiano, G. 2004, ApJ, 613, 933
Kim, D.-W. et al. 2008, ApJ, 688, 931
Kim, D.-W., 2011, in *Hot Interstellar Matter in Elliptical Galaxies*, eds. D.-W. Kim and S. Pellegrini,
    Astrophysics and Space Science Library, Springer (NY), in press
Kim, D.-W. & Fabbiano, G. 2010, ApJ, 721, 1523
Koleva, M., Prugniel, P., de Rijcke, S., and Zeilinger, W. 2011 arxiv/1105.4809.
Kuntschner, H., et al. 2010. arxiv /1006.1574.
Loewenstein, M., & Davis, D. S. 2010, ApJ, 716, 384
Maeda, K. et al. 2010, ApJ, 712, 624
Matsushita, K., Ohashi, T., & Makishima, K. 2000, PASJ 52, 685
Nagataki, S., & Sato, K 1998, ApJ, 504, 629
Pipino, A., Kawata, D., Gibson, B. K, & Matteucci, F. 2005, AA, 434, 553
Pipino A., D'Ercole A., Matteucci F., 2008, A&A, 484, 679
Pipino, A. 2011, in *Hot Interstellar Matter in Elliptical Galaxies*, eds. D.-W. Kim and S. Pellegrini,
    Astrophysics and Space Science Library, Springer (NY), in press
Pipino, A. & Matteucci, F. 2011, AA, 530, 98
Richings, A. J. et al. 2010, ApJ, 723, 1375
Sanders, J. S., & Fabian, A. C. 2002, MNRAS, 331, 273
Sansom, A. E. et al. 2006, MN, 370, 1541

Schweizer, F. 2003, in ASP Conf. Ser. 296, New Horizons in Globular Cluster Astronomy, ed. G. Piotto (San Francisco, CA: ASP), 467

Serra, P., and Trager, S. C., 2007, MNRAS, 374, 769

Serra, P., and Oosterloo, T. A. 2010, MNRAS, 401, L29

Smith, R. K. et al. 2001, ApJ, 556, L91

Tal, T., van Dokkum, P. G., Nelan, J. and Bezanson, R. 2009. AJ, 138, 1417

Terlevich, A. I. & Forbes, D. A 2002, MN 330, 547

Thomas, D. et al. 2005, ApJ, 621, 673

Tonry, J. L., et al. 2001, ApJ, 546, 681

Trager, S. C. et al. 2000, AJ, 120, 165

Trager, S. C. & Somerville, R. S. 2009 MNRAS, 395, 608

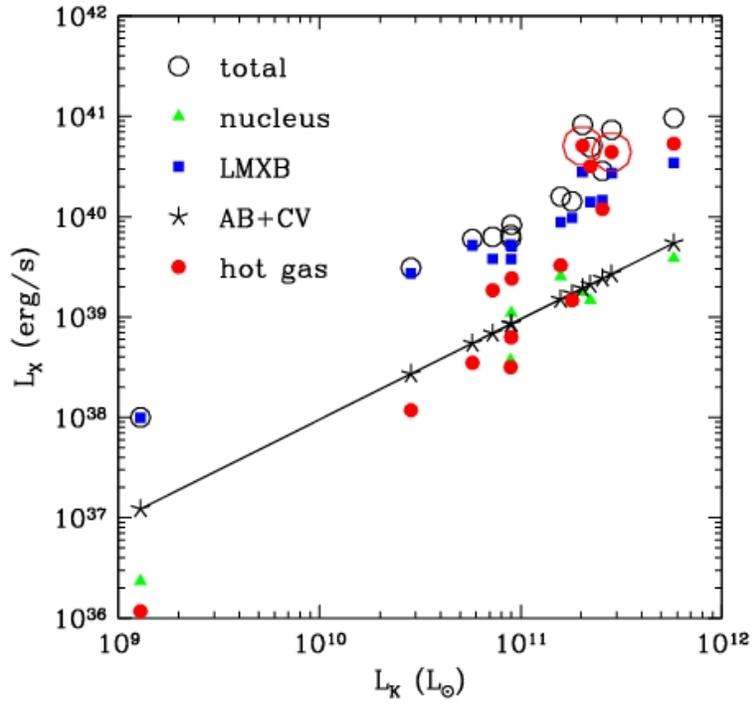

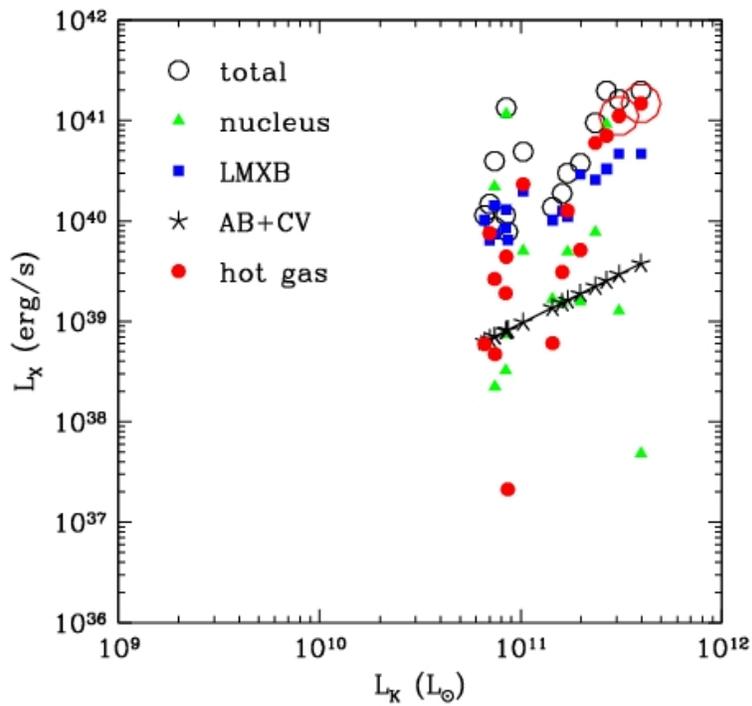

**Figure 1**. X-ray luminosity ($L_X$) is plotted against the K-band luminosity ($L_K$) of (a) young and (b) old early type galaxies in BKF sample. The X-ray luminosity from different components are marked by different symbols: black open circles for the total $L_X$, red filled circles for $L_X$(gas), blue filled circles for $L_X$(LMXB), green triangles for $L_X$(nucleus), and black stars for $L_X$(AB+CV). We refer to BKF for how to estimate the X-ray luminosities from different components. Our sample galaxies are also marked by big circles at their $L_X$(gas).

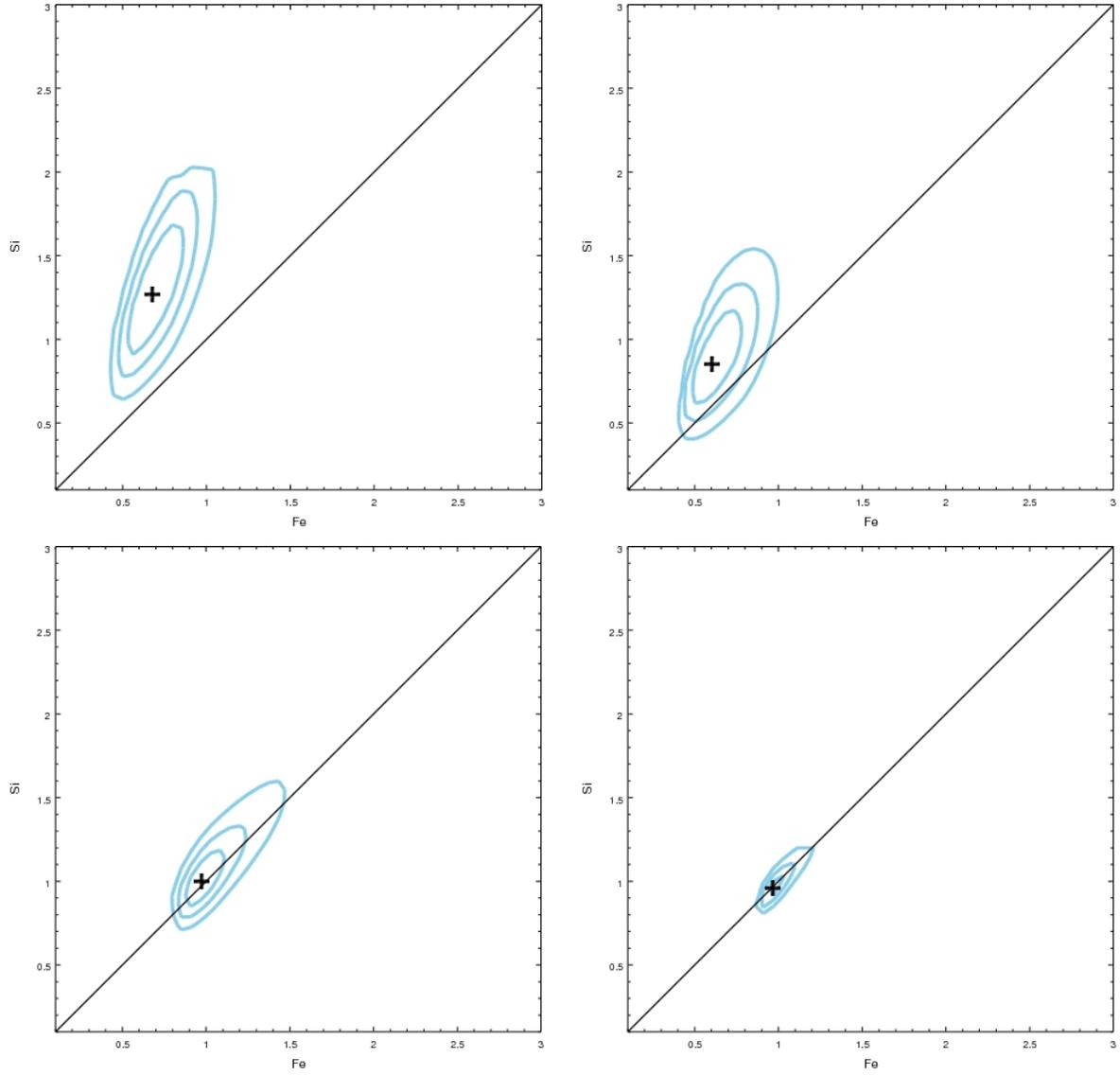

**Figure 2**. Confidence contours of *Si* and *Fe* abundances in the solar unit for (a) NGC 720, (b) NGC 3923, (c) NGC 4472 and (d) NGC 4649. Three contours in each plot are at 39%, 68% and 90% confidence levels

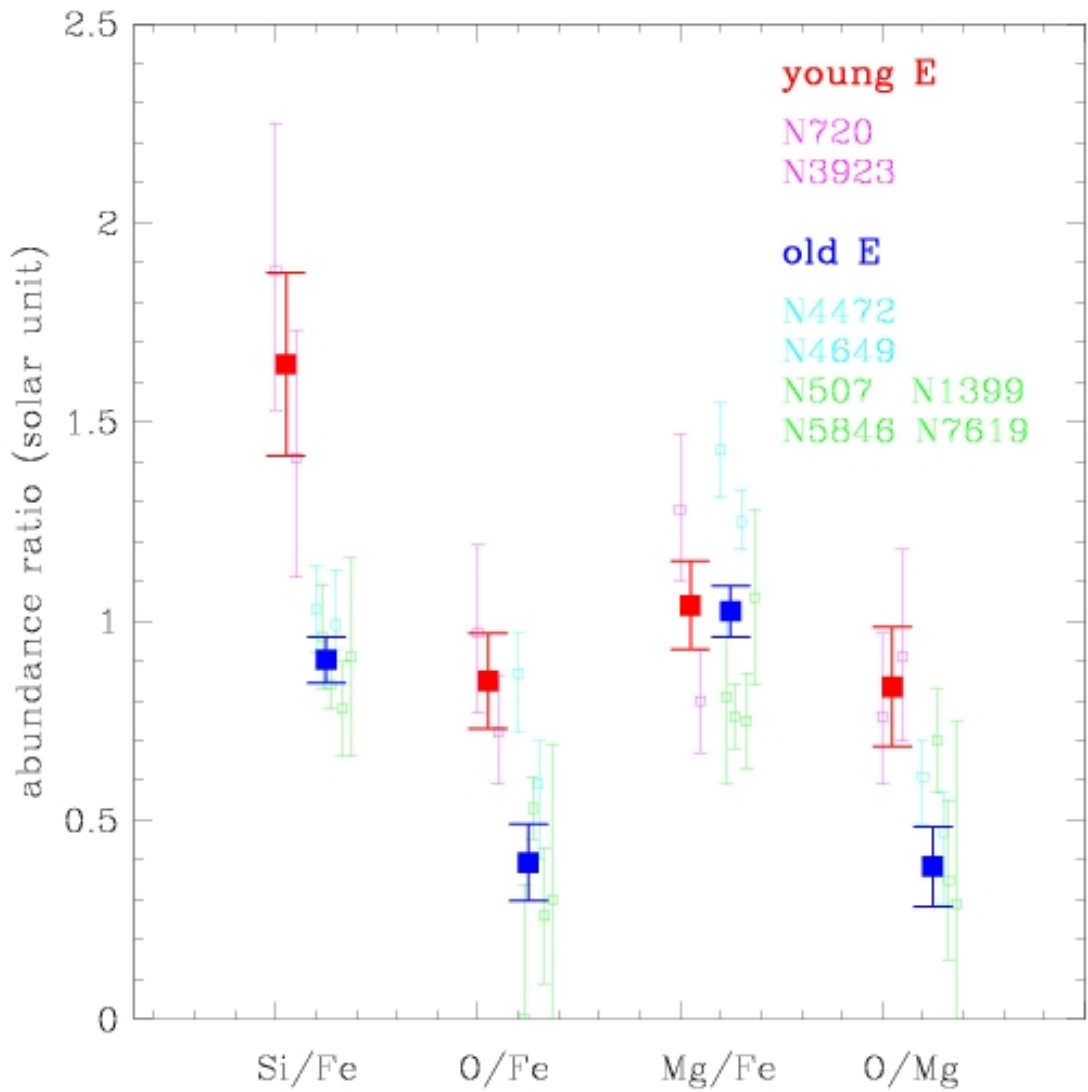

**Figure 3**. Comparison of the abundance ratios (in solar unit) measured with X-ray spectra extracted from the central regions (r < 30") of two young elliptical galaxies and six old elliptical galaxies. The data for four additional old elliptical galaxies are taken from Humphrey and Buote (2006).

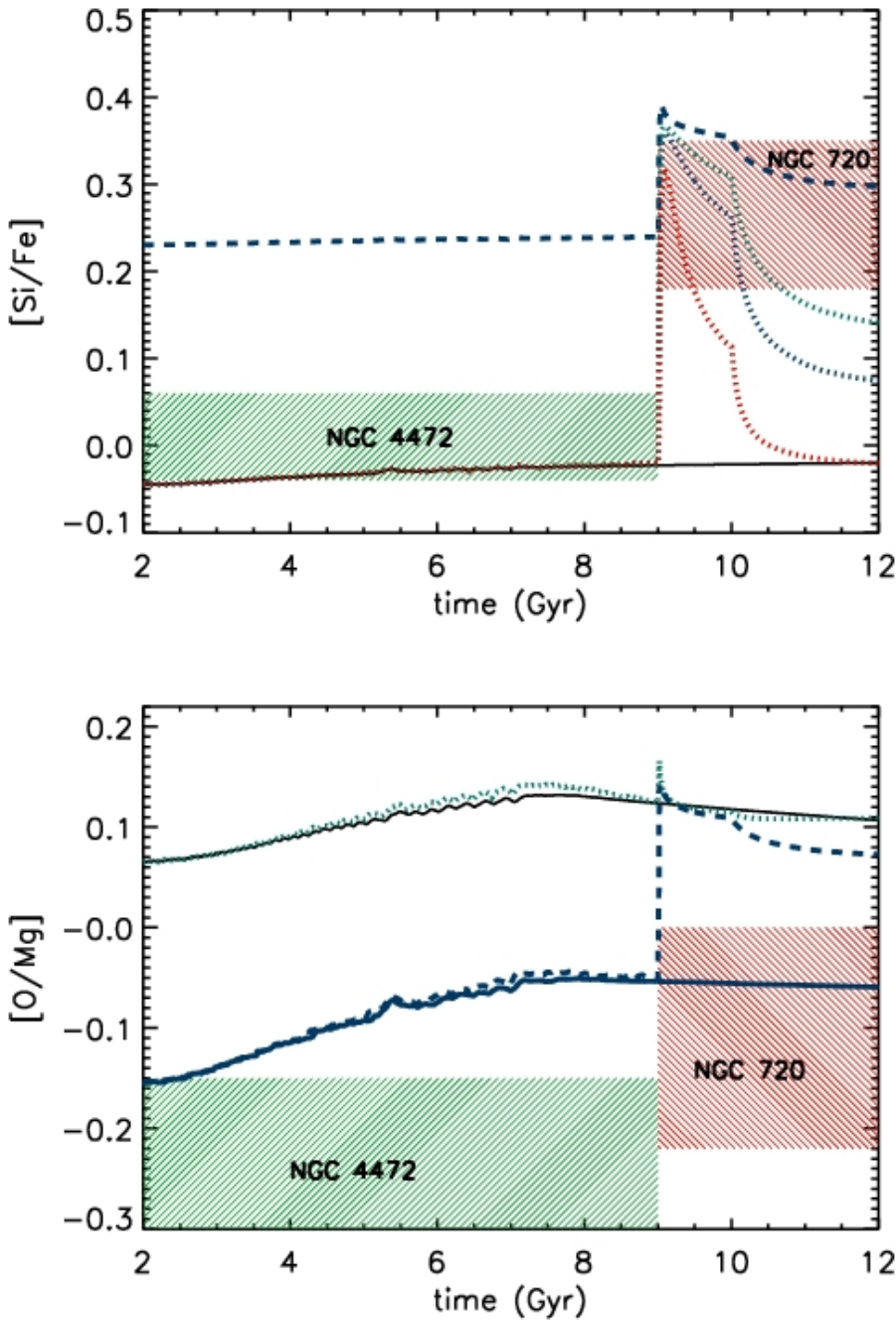

**Figure 4**. (a) The *Si/Fe* abundance ratio in the ISM predicted by chemical evolution models to illustrate the effect of the 2$^{nd}$ star formation. The black solid line is for the base model, case 2a in Pipino & Matteucci (2011). The 2$^{nd}$ star formation is ignited at an arbitrary age of 9 Gyr for 1 Gyr (red dotted). The SN Ia rate is further reduced to match with the observational estimate in clusters (blue dotted) and in the field (green dotted). The blue dashed line indicates the result with the new SN 1a yields taken from the C-DDT model (Maeda et al. 2010) which corresponds to case 3a in Pipino & Matteucci (2011). Also plotted are the observed ranges for NGC 720 (red shaded) and NGC 4472 (blue shaded). (b) same as (a), but for the *O/Mg* abundance ratio. The green dotted and blue dash lines show the results of the additional star formation from the base models of case 2a (black solid) and case 3a (blue solid), respectively.